\documentclass[aps,prb,amsmath,amssymb,twocolumn,10pt,superscriptaddress]{revtex4-1}

  \usepackage{times,txfonts}
  \usepackage{gensymb}
  \usepackage{siunitx}
  \usepackage{dcolumn}
  \usepackage{multirow}
  \usepackage[normalem]{ulem}
  \usepackage{times}
  \usepackage[utf8]{inputenc}
  \usepackage[T1]{fontenc}
  
  \usepackage[]{graphicx}
\usepackage[dvipsnames]{xcolor}
\usepackage{tabularx}

\usepackage{amsfonts}
\usepackage{amsmath}
\usepackage{amssymb}
\usepackage{bm}
\usepackage{enumerate}
\usepackage[breaklinks,colorlinks=true,citecolor=blue]{hyperref}

\begin{document}
\title{Field-controlled quantum anomalous Hall effect in electron-doped CrSiTe$_{ 3 }$ monolayer: a first-principles prediction}

\author{Sungmo Kang}
\affiliation{Center for Theoretical Physics, Department of Physics and Astronomy, Seoul National University, Seoul 08826, Republic of Korea}
\author{Seungjin Kang}
\affiliation{Center for Theoretical Physics, Department of Physics and Astronomy, Seoul National University, Seoul 08826, Republic of Korea}
\author{Heung-Sik Kim}
\email{heungsikim@kangwon.ac.kr}
\affiliation{Department of Physics and Institute of Quantum Convergence Technology, Kangwon National University, Chuncheon 24311, Republic of Korea}
\author{Jaejun Yu}
\email{jyu@snu.ac.kr}
\affiliation{Center for Theoretical Physics, Department of Physics and Astronomy, Seoul National University, Seoul 08826, Republic of Korea}

\begin{abstract}
We report Chern insulating phases emerging from a single layer of layered chalcogenide CrSiTe$_{3}$, a transition metal trichacogenides (TMTC) material, in the presence of charge doping. Due to strong hybridization with Te $p$ orbitals, the spin-orbit coupling effect opens a finite band gap, leading to a nontrivial topology of the Cr $e_{\mathrm{g}}$ conduction band manifold with higher Chern numbers. Our calculations show that quantum anomalous Hall effects can be realized by adding one electron in a formula unit cell of Cr$_{2}$Si$_{2}$Te$_{6}$, equivalent to electron doping by 2.36$\times$10$^{14}$ cm$^{-2}$ carrier density. Furthermore, the doping-induced anomalous Hall conductivity can be controlled by an external magnetic field via spin-orientation-dependent tuning of the spin-orbit coupling. In addition, we find distinct quantum anomalous Hall phases employing tight-binding model analysis, suggesting that CrSiTe$_{3}$ can be a fascinating new platform to realize Chern insulating systems with higher Chern numbers.
\end{abstract}


\maketitle

\section{INTRODUCTION}
\label{sec:introd}
Since the discovery of graphene, two-dimensional (2D) materials have been one of the most intensively studied systems owing to their intriguing physical properties and the chance of applications to atomistic low-power devices\cite{geim2007graphene,PhysRevB.76.073103,hBN2010natnano}.
Transition metal chalcogenides (TMC) with layered structures is a representative example of such 2D materials\cite{wang2012mos2,mak2012mos2}. One intensively pursued subject recently in TMC is magnetism originating from the presence of localized magnetic moments in $d$ orbitals at TM ions, where hybridization between TM and chalcogen anions give rise to interesting magnetic behaviours in $X$PS$_3$ ($X$ = Mn, Fe, Ni) and Cr$B$Te$_3$ ($B$ = Si, Ge)\cite{PhysRevB.91.235425,doi:10.1021/acs.nanolett.6b03052,li2014crxte,PhysRevB.92.224408,doi:10.1021/acs.nanolett.9b05165}. Systems like Cr$B$Te$_3$ or CrI$_3$ are reported to show Ising-type magnetism and exhibit magnetic orders down to single-layer limit\cite{PhysRevB.92.035407} at finite temperatures\cite{gong2017cgt,Huang2017cri3}, defying the Mermin-Wagner theorem that prohibits long-range order in systems with d ${\leq}$ 2 dimensions via thermal fluctuations\cite{PhysRevLett.17.1133}. Successful exfoliations of atomistically-thin CrI$_3$ and CrGeTe$_3$ layers presents promising magnetic 2D system with potential device applications\cite{gong2017cgt,Huang2017cri3}.

The presence of robust magnetism in two-dimensional system like Cr$B$Te$_3$ becomes even more interesting in the study of topological properties\cite{PhysRevLett.61.2015,PhysRevB.90.121103,chen2017intrinsic,Niu2017QuantumSH,PhysRevX.4.011010,PhysRevB.85.045445,PhysRevB.94.125134,PhysRevB.93.184306,PhysRevB.98.155148,Baidya2019}, where TMC materials has been recently considered as one of good platform of researching topological phases of matters\cite{PhysRevB.97.035125,PhysRevLett.117.257201} and magnetism can be employed as a control knob to tune topological properties in such systems\cite{PhysRevB.103.195152,PhysRevB.104.054422,PhysRevB.104.165424}. Specifically, quantum anomalous Hall effect (QAHE) can be observed in materials with magnetic ordering and sizable band gap, showing quantized integer Hall conductivity in the absence of external magnetic fields. Chern insulator is one of topological phases of matter showing QAHE characterized by \textbf{Z} topological invariant, i.e. Chern numbers,  unlike time-reversal-symmetric conventional topological insulators classified by \textbf{Z}$_2$-invariants such as Bi$_2$Se$_3$, Bi$_2$Te$_3$, Sb$_2$Te$_3$\cite{nphys1270} or HgTe\cite{PhysRevB.99.115411}.
Recent studies suggest Chern insulating phases with high Chern numbers and wide bandgap energy\cite{PhysRevB.97.035125,PhysRevLett.117.257201} can be applied for low-power dissipationless device applications, but exploring suitable candidates is a nontrivial task. CrSiTe$_{3}$, originally reported to be a 2D ferromagnetic semiconductors with trivial band topology\cite{chen2015strain,lin2016ultrathin,PhysRevB.92.144404,li2014crxte},
can be an ideal candidate in the presence of charge doping because of the robust magnetism and the insulating behavior down to the single-layer limit, in addition, due to the presence of strong SOC from Te sites. 

In this work, by performing first-principles density-functional theory (DFT) calculations, we find Chern insulating phases in CrSiTe$_{3}$ monolayer in the presence of electron doping. We find crossing points within the Cr ${e}_{\mathrm{g}}$ conduction band manifolds, where the crossings are removed as spin-orbit-coupling (SOC) is introduced and leading to topologically nontrivial bands with Chern number up to 8. We also find quantized anomalous Hall conductivity (AHC) at one- and two-electron-doping per unit cell, which can be further varied as spin orientation angle changes. Additionally, it is found that the magnetic anisotropy energy (MAE) is suppressed as electron doping is introduced via electrostatic gating, consistently with recent findings\cite{nelecton2520,doi:10.1021/acs.nanolett.9b03815,https://doi.org/10.1002/adma.202008586}, which makes tuning of AHC via external magnetic field feasible. Finally, we construct a tight-binding (TB) hamiltonian for a deeper understanding of our DFT results and to further pursue the possibility of realizing distinct Chern insulating phases via external stimuli such as strain. 
Our result reveals that single-layers of magnetic CrSiTe$_3$ and its structural siblings are promising platforms to realize Chern insulator materials with high Chern numbers. 

\begin{figure}
  \includegraphics[width=1.0\linewidth]{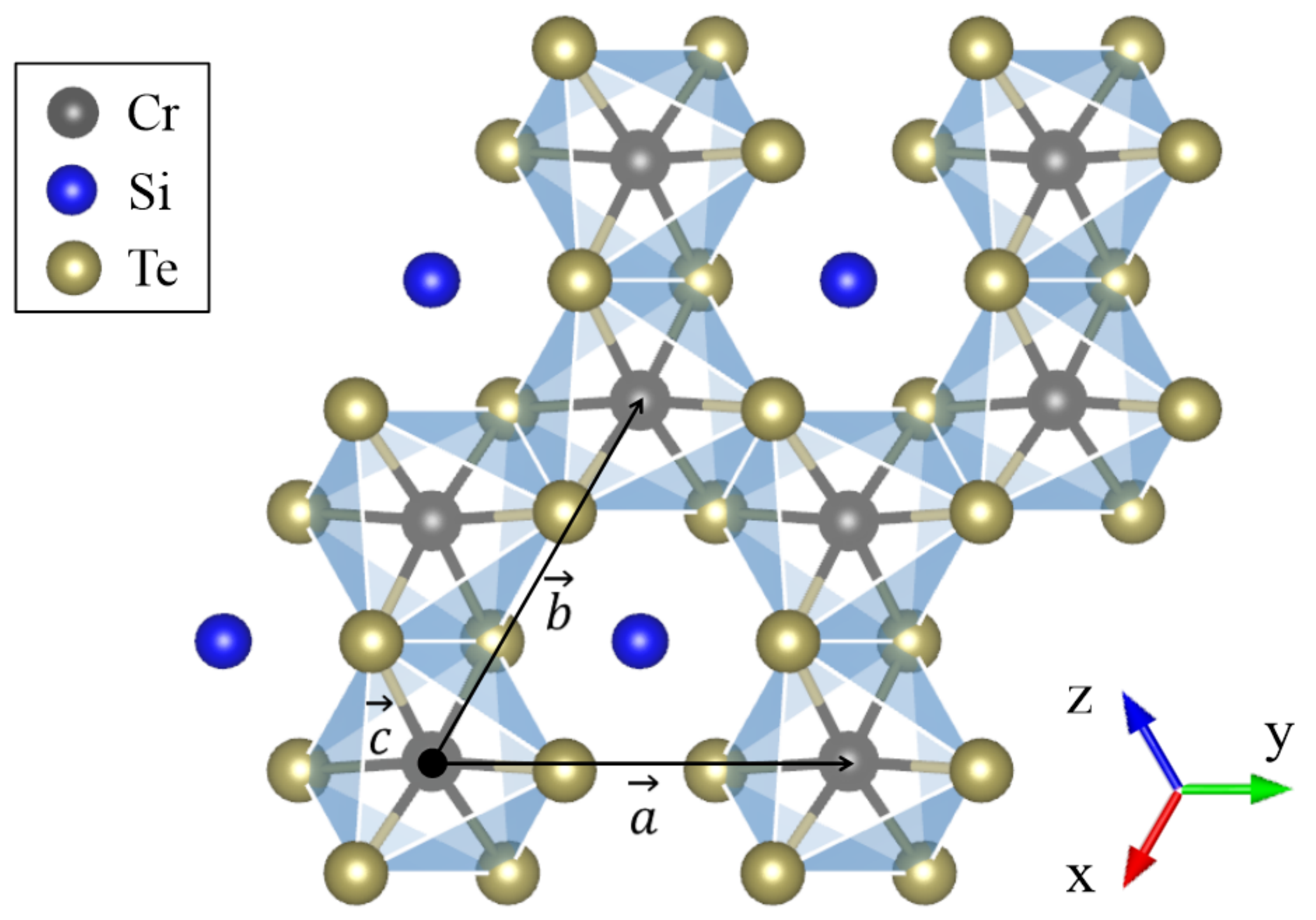}
  \caption{\textbf{Crystal structure of monolayer CrSiTe$_{ 3 }$ (top view).} The unit cell features the Cr$_{2}$Si$_{2}$Te$_{6}$ formula unit. Edge sharing CrTe$_{6}$ octahedron consists of Cr atom (black sphere) and surrounding six Te atoms (yellow sphere). Dimerized two Si atoms (blue sphere) are sitting at center of Cr honeycomb lattice. \textbf{a},\textbf{b},\textbf{c} is unit vector of hexagonal unit cell, where \textbf{x},\textbf{y},\textbf{z} indicates local axis.}
  \label{fig:structure}
 \end{figure}

\begin{figure}
	\includegraphics[width=1.0\linewidth]{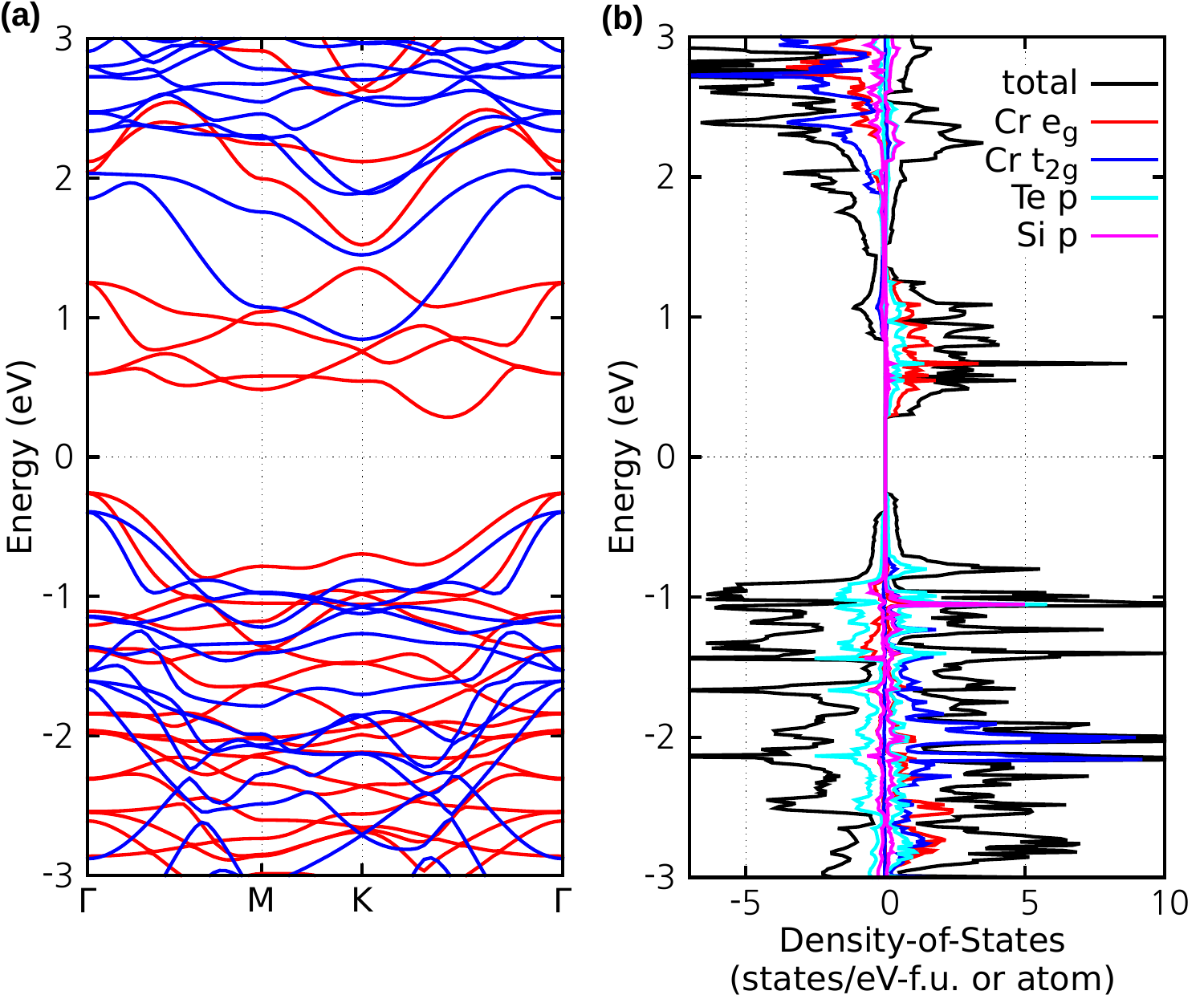}
    \caption{\textbf{Band structure and pDOS plot}. \textbf{(a)} Band structure with majority spin (red line) and minority spin channel (blue line), respectively. \textbf{(b)} Total density of states (black line), projected density of states of Cr ${e}_{\mathrm{g}}$ orbital (red line) and ${t}_{\mathrm{2g}}$ orbital (blue line), Si ${p}$ orbital (purple line) and Te ${p}$ orbital (sky blue line), respectively. In the pDOS plot, plus and minus signs correspond to the majority and minority spin components, respectively. Fermi level is set to be zero for both panels.}
	\label{fig:band-pdos}
\end{figure}

\begin{figure*}
	\includegraphics[width=1.0\linewidth]{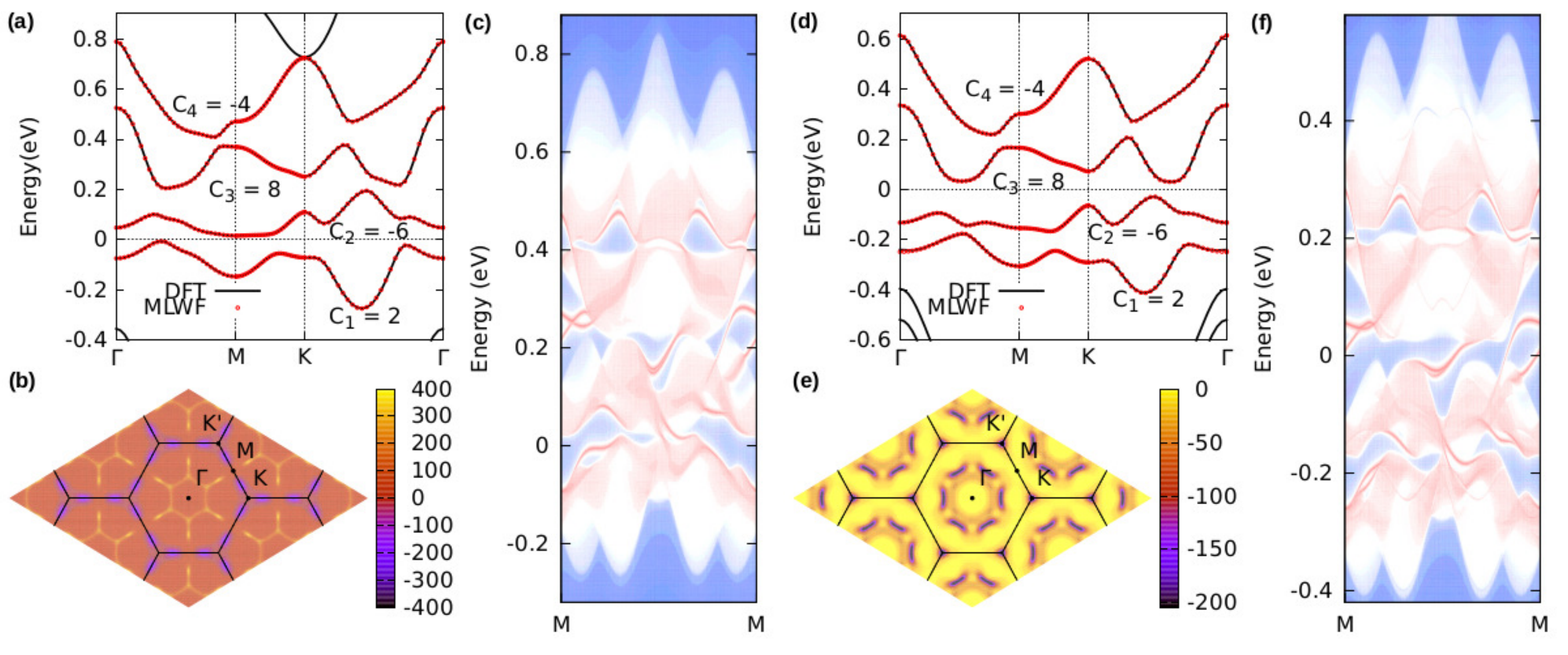}
	\caption{\textbf{Chern number, Berry curvature and chiral edge modes.} Cr ${e}_{\mathrm{g}}$ bands manifold calculated by using DFT (black line) and Wannier function interpolation (red circle) under \textbf{(a)} one electron doping and \textbf{(d)} two electron doping per Cr$_{2}$Si$_{2}$Te$_{6}$ formula unit, respectively. Chern numbers for each band are also remarked. Berry curvature plot of the occupied Cr ${e}_{\mathrm{g}}$ band in momentum space under \textbf{(b)} one electron doping and \textbf{(e)} two electron doping per formula unit, respectively. Unit of Berry curvature is 1/{\AA}$^{2}$. Plot of chiral edge states along armchair direction under \textbf{(c)} one electron doping and \textbf{(f)} two electron doping per formula unit, respectively. Red lines indicate edge state components. Pink regions are bulk states, while the blue area corresponds to vacuum. All panels are drawn using data calculated by introducing SOC effect with ferromagnetic ordering to out-of-plane direction.}
	\label{fig:berry-chern}
\end{figure*}

\section{\label{sec:results}RESULT}
\subsection{\label{sub:fm-ground} Electronic and magnetic properties of stoichimetric CrSiTe$_{3}$}

CrSiTe$_{3}$ is one of ${MAX}_{\mathrm{3}}$-type TMTC compounds. It has a $R\overline{3}$-type stacking of neighboring CrSiTe$_{3}$ monolayers, where neighboring layers are coupled by vdW interaction. Crystal structure of a CrSiTe$_{3}$ monolayer is depicted in Fig.~\ref{fig:structure}, showing edge-sharing honeycomb network of CrTe$_{6}$ octahedra and Si-dimers located at the centers of Cr-hexagons.
In undoped CrSiTe$_{ 3 }$, Cr cations are in the $d^3$ (Cr$^{3+}$) high-spin configuration ($S=3/2$) with fully occupied ${t}_{\mathrm{2g}}$ orbitals. In addition, strong $dp{\sigma}$ hybridization between Cr $d$- and Te $p$-orbitals gives rise to additional magnetic moment contributions, so that the magnitude of Cr spin moment is 3.87 $\mu_{\rm B}$\cite{kang2019cstcgt}. The ferromagnetic ground state can be attributed to the Goodenough-Kanamori-Anderson (GKA) superexchange mechanism\cite{GOODENOUGH1958287,PhysRev.100.564}. Our results are in agreement with previous reports\cite{casto2015strong,li2014crxte,chen2015strain} where choice of Hubbard U parameters is crucial for determining magnetic ground state of CrSiTe$_{3}$\cite{kang2019cstcgt}.

When electron doping is introduced, orbital occupation of CrSiTe$_{3}$ changes so that magnitude of the magnetic moment of Cr $d$ orbitals, as well as lattice constant, is also affected. With two-electron doping per formula unit cell, Cr ${e}_{\mathrm{g}}$ orbital of spin majority channel becomes half-filled so that the high-spin Cr$^{2+}$ ($S=2$) configuration is obtained (additional moment of 0.51 ${\mu}_{B}$ further induced due to $dp{\sigma}$-hybridization). With this electron doping, the in-plane lattice parameter $a$ expands to be 7.26 \AA, 5.7\% larger than that of undoped case\cite{PhysRevB.91.235425,kang2019cstcgt,chen2015strain}. We check that the ferromagnetic ground state with insulating properties and also topological phases remain unchanged within the Hubbard U parameters and lattice constant range of 0.5 eV ${\leq}$ $U$ ${\leq}$ 1.5 eV and 6.87 \AA\ ${\leq}$ $a$ ${\leq}$ 7.26 \AA, respectively. Therefore, hereafter we choose the lattice constant and Hubbard $U$ parameters to be 7.0 \AA\ and 1.5 eV, respectively, as a representative case.

\subsection{Band crossings in CrSiTe$_{3}$ $\bf\emph{e}_{g}$ conduction bands}
\label{sub:e_g-bands}
To study magnetic and topological properties of electron-doped monolayer CrSiTe$_{3}$ in the FM state, we first focus on the electronic structure of undoped CrSiTe$_{3}$. 
Fig.~\ref{fig:band-pdos} describes band structures and projected density-of-states (pDOS) of ferromagnetic single layer CrSiTe$_{3}$.  Majority- and minority-spin components show exchange splitting where the minority-spin parts have large band gap energy compared to the majority-spin parts. Right above the Fermi level, four majority-spin bands consisting of Cr ${e}_{\mathrm{g}}$- and Te ${p}$-orbitals exist and are separated from other bands (except Cr ${t}_{\mathrm{2g}}$ bands in the minority-spin channel). Interestingly, band crossings are found within the ${e}_{\mathrm{g}}$ bands at $\Gamma$, $K$ points and on the $\Gamma$-M, $\Gamma$-K high-symmetry lines. This observation suggests the possibility for the monolayer CrSiTe$_{3}$ to host nontrivial band topology in the presence of electron doping and gap opening via SOC, as discussed in the following section.

\subsection{Berry curvature plot, Chern numbers and chiral edge states}
\label{berry-plot}
Fig.~\ref{fig:berry-chern} shows magnified Cr ${e}_{\mathrm{g}}$ band structures with Chern numbers assigned to each band, plots of Berry curvature in the momentum space, and edge spectra in the presence of integer electron doping, SOC, and out-of-plane ferromagnetic spin orientation. We consider the cases of one and two electron doping per formula unit cell, where both systems are insulating as shown in Fig.~\ref{fig:berry-chern}(a) and (d). For each band, we find unusually high Chern numbers of up to 8, with about 10 meV order of SOC-induced band gap mainly originating from Te ${p}$ orbital contributions. 

Fig.~\ref{fig:berry-chern}(b) and (e) present Berry curvature of occupied Cr ${e}_{\mathrm{g}}$ band in the presence of one and two electron doping per formula unit cell, respectively. In the presence of one electron doping per unit cell (Fig.~\ref{fig:berry-chern}(b)), peaks of positive Berry curvature for the lowest Cr ${e}_{\mathrm{g}}$ band appear around $\Gamma$ point and on the $\Gamma$ - $M$ lines, while negative peaks appear close to $K$ and $K'$ points, yielding the Chern number of 2 for the lowest $e_{\rm g}$ band. In the case of half-filled $e_{\rm g}$ (Fig.~\ref{fig:berry-chern}(e)), two peaks of negative Berry curvature are located at $K$ and $K'$ points. In addition, six negative peaks are located between six $\Gamma$ - $K$ lines, resulting in a total Chern number of $-$4. Fig.~\ref{fig:berry-chern}(c) and (f) show edge spectra from the one and two-electron doping (per formula unit cell) results, respectively. In agreement with the Chern number calculation results, one can find two and four chiral edge states near the Fermi level in Fig.~\ref{fig:berry-chern}(c) and (f), respectively. The emergence of multiple band crossings in Cr ${e}_{\mathrm{g}}$ bands, as shown in Fig.~\ref{fig:band-pdos}(a), gives rise to nontrivial topology with high Chern numbers. Explanation of microscopic origin of multiple Dirac cones in previous theoretical work\cite{PhysRevB.97.035125} can also be applied in our case, where hopping integrals calculation results of CrSiTe$_{ 3 }$ are listed in Supplementary information section 1. In addition, descriptions for band gap closing at the midpoint of the $\Gamma$ - $K$ line are discussed. See more details in Supplementary information section 2.

\begin{figure}
	\includegraphics[width=1.0\linewidth]{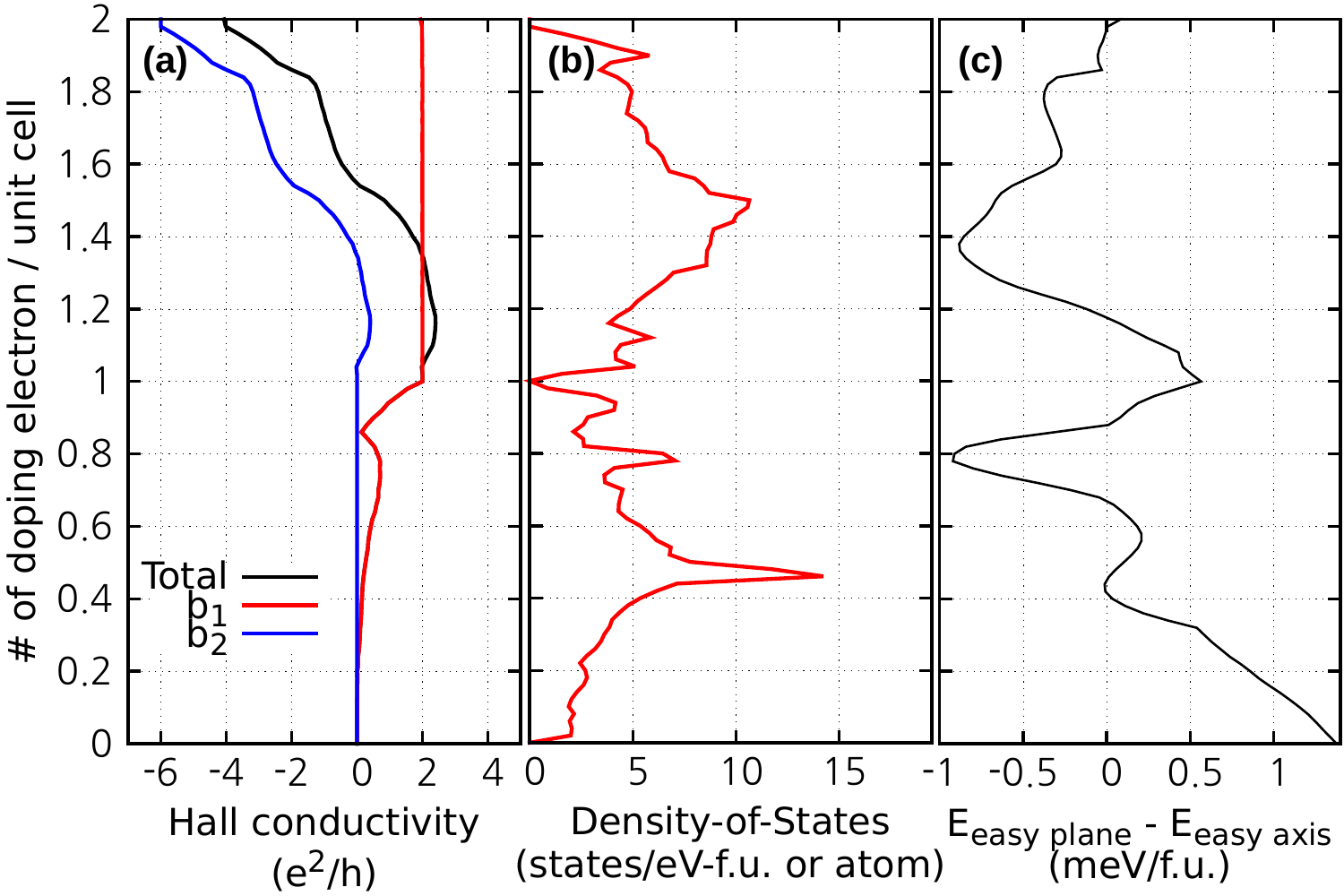}
	\caption{\textbf{Electron doping dependent AHC, DOS and MAE.} \textbf{(a)} AHC contributions for each band of Cr $\bf\emph{e}_{g}$ band manifolds from the lowest band `b1' (red line), the second-lowest band `b2' (blue line), and the sum of two bands `Total' (black line), respectively. \textbf{(b)} The total density-of-states of Cr ${e}_{\mathrm{g}}$ bands. Note that the results of panel \textbf{(a)} and \textbf{(b)} are obtained in the presence of the out-of-plane spin orientation. \textbf{(c)} The energy difference between easy plane and easy axis configurations of ferromagnetic monolayer CrSiTe$_{ 3 }$.}
	\label{fig:band-ahc}
\end{figure}

\subsection{AHC under electron doping}
\label{sub:anomalous Hall conductivity}
We further examine the AHC of Cr ${e}_{\mathrm{g}}$ bands manifold, the total density-of-states of CrSiTe$_{ 3 }$ and MAE illustrated in Fig.~\ref{fig:band-ahc}. Total AHC shows a quantized plateau at an integer number of electron doping, consistent again with the Chern number calculation results in Fig.~\ref{fig:berry-chern}(a) and (d). In addition, the total density-of-states vanishes at one- and two-electron doping as depicted in Fig.~\ref{fig:band-ahc}(b). Therefore, Cr ${e}_{\mathrm{g}}$ bands are separated from each other together with nonzero Chern numbers so that Chern insulating phases will be realized under one- or two-electron doping in the formula unit cell. 

\subsection{Magnetic anisotropy and its doping dependence}
\label{sub:dipole}
So far, our DFT calculations with SOC effect consider spin aligned to the out-of-plane direction. However, this out-of-plane spin configuration can be suppressed or become unstable as the electron doping is introduced because magnetic anisotropy may depend on $e_{\rm g}$ occupation. Fig.~\ref{fig:band-ahc}(c) shows MAE as a function of electron doping concentrations, which oscillate between easy-plane and easy-axis anisotropies. For example, easy-axis anisotropy occurs at one electron doping, while two-electron doping per formula unit cell favors easy-plane anisotropy. It is worth mentioning that MAE may vanish as doping is introduced, which may enable tuning the direction of FM moments and the resulting electronic structure via external magnetic fields.

\begin{figure}
	\includegraphics[width=0.9\linewidth]{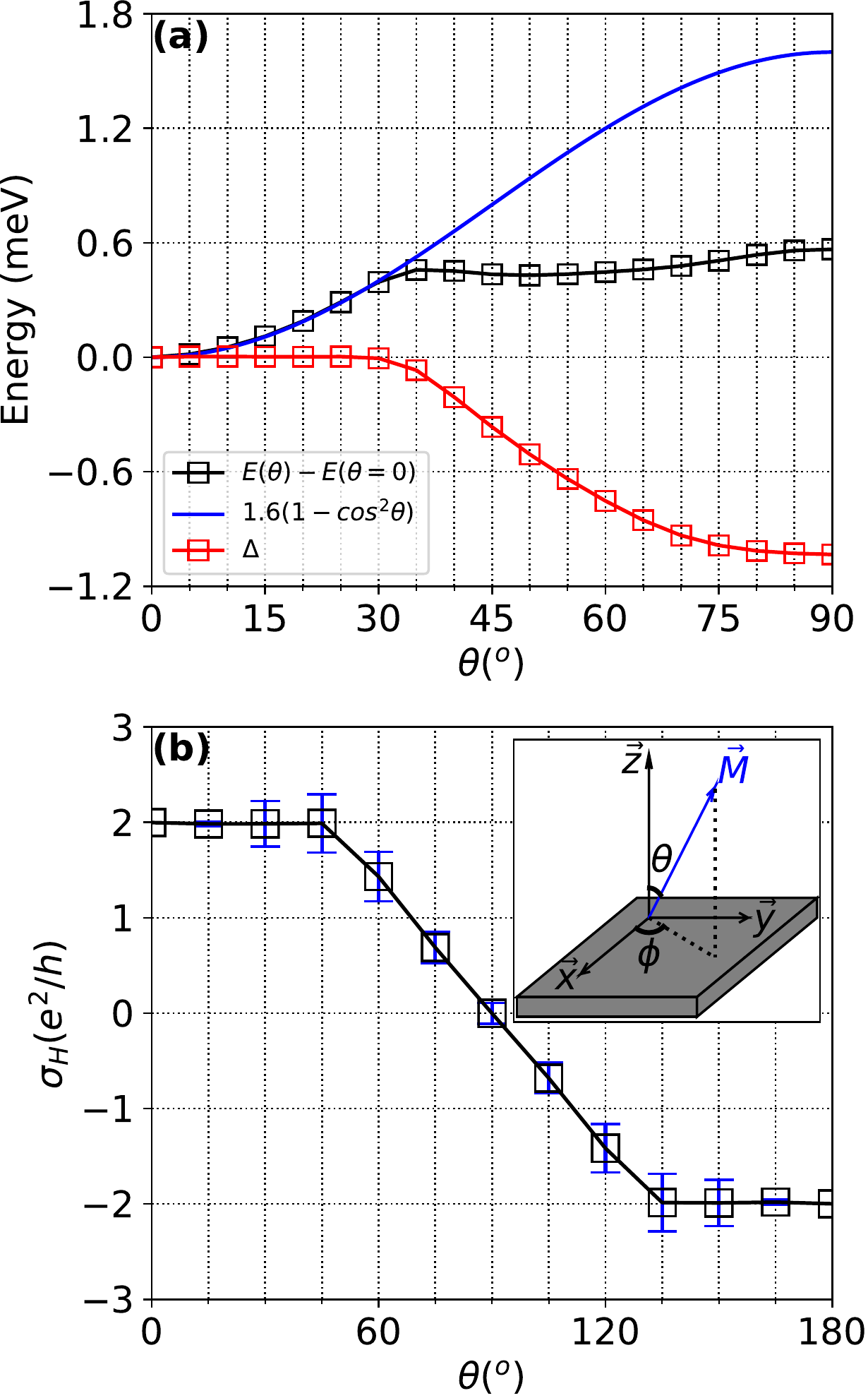}
	\caption{\textbf{Total energy and AHC as a function of spin-orientation angle} \textbf{(a)} Total energy at spin-orientation angle ${\theta}$ (black line and marker), where total energy of ${\theta}$ = 0 is set to be zero. By interpolating the calculated total energy of 0 ${\leq}$ ${\theta}$ ${\leq}$ 30, we obtain fitting function indicated as blue line. Difference between total energy and the fitting function are depicted as red line and marker. \textbf{(b)} Azimuthal spin angle (${\phi}$) averaged AHC as a function of polar angle ({$\theta$}). Mean value (black line with square marker) and standard deviation (blue errorbar) of AHC for each {$\theta$} are remarked. Upper right panel shows definition of polar angle ({$\theta$}) and azimuthal spin angle (${\phi}$) of magnetization vector (blue arrow) and CrSiTe$_{ 3 }$ monolayer as gray slab.}
	\label{fig:AHC_MAE_theta}
\end{figure}

\begin{table}[h]
\renewcommand{\arraystretch}{1.1}
\renewcommand{\tabcolsep}{2.0mm}
\centering
\begin{tabular}{c c c c}
 \hline
 \# of doping  & magnetic & MAE & D-MAE\\
 electron ($\Delta$n) & moment (${\mu}_{B}$) & (meV/f.u.) & (meV/f.u.)\\
 \hline\hline
 0.0 & 3.97 & 1.378 & -0.244 \\
 0.5 & 4.10 & 0.101 & -0.260\\
 1.0 & 4.23 & 0.566 & -0.277 \\
 1.5 & 4.37 & -0.665 & -0.295\\
 2.0 & 4.51 & 0.085 & -0.314 \\
 \hline
\end{tabular}
\caption{Magnetic moment and anisotropy energies for each number of doping electrons in the formula unit cell. MAE is defined by the energy difference between the easy axis and easy plane spin configuration, estimated by DFT calculations. Magnetic dipolar anisotropy energy (D-MAE) is magnetic dipole-dipole interaction energy. The (+) sign indicates the preference for out-of-plane spin orientation. All calculations are performed with a fixed lattice constant of 7.0 \AA.}
\label{table:MAE}
\end{table}

\begin{figure*}
	\includegraphics[width=0.9\linewidth]{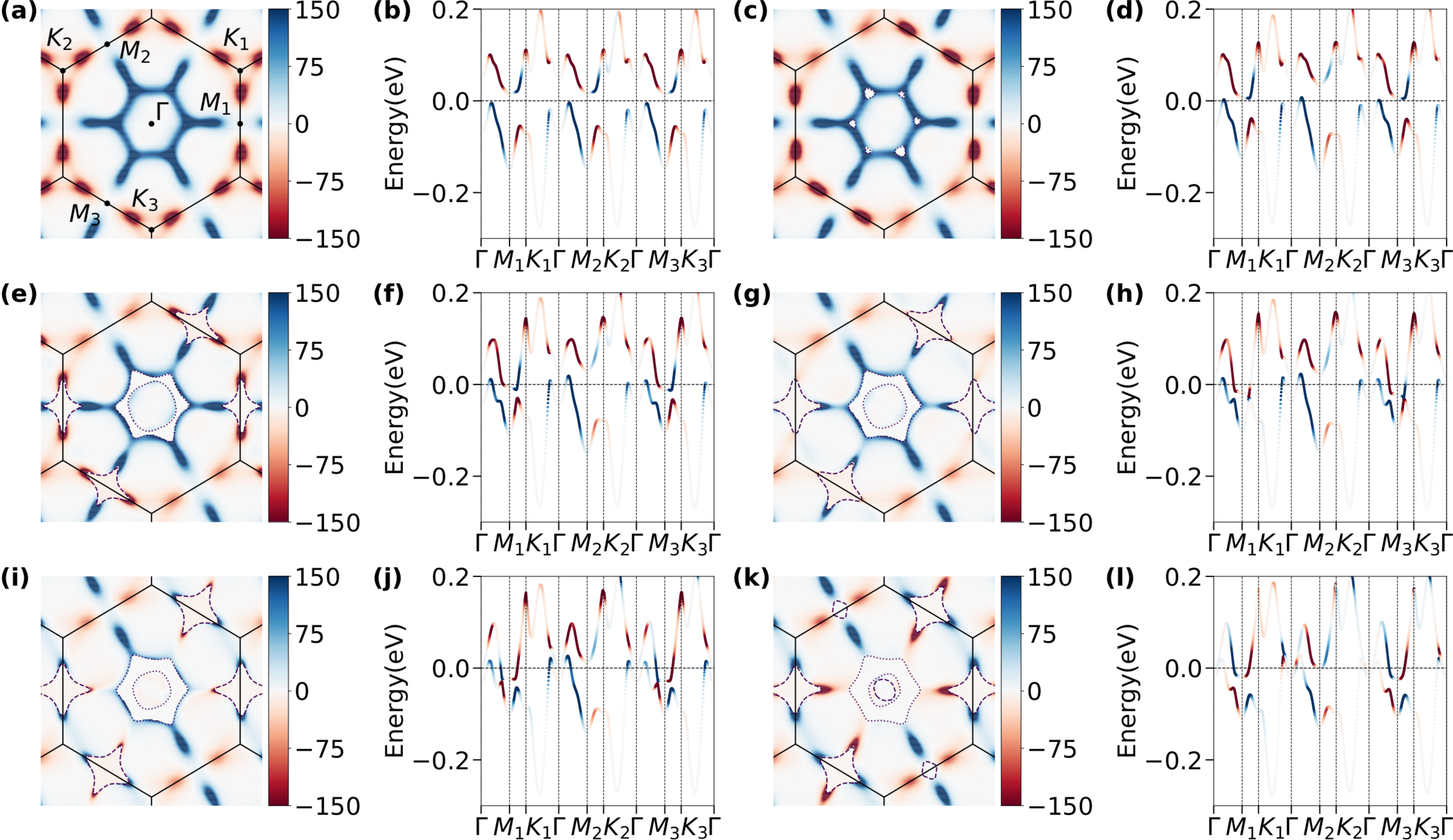}
	\caption{\textbf{Spin orientation angle dependent Berry curvature plot} Berry curvatures of occupied Cr ${e}_{\mathrm{g}}$ states in the momentum space (unit in ${\AA}^2$) for spin-orientation angle(${\theta}$) \textbf{(a)} ${\theta}$ = 0$^{\circ}$, \textbf{(c)} ${\theta}$ = 30$^{\circ}$, \textbf{(e)} ${\theta}$ = 45$^{\circ}$, \textbf{(g)} ${\theta}$ = 60$^{\circ}$, \textbf{(i)} ${\theta}$ = 75$^{\circ}$ and \textbf{(k)} ${\theta}$ = 90$^{\circ}$, respectively. Berry-curvature-projected band structures of Cr ${e}_{\mathrm{g}}$ bands near Fermi level for \textbf{(b)} ${\theta}$ = 0$^{\circ}$, \textbf{(d)} ${\theta}$ = 30$^{\circ}$, \textbf{(f)} ${\theta}$ = 45$^{\circ}$, \textbf{(h)} ${\theta}$ = 60$^{\circ}$, \textbf{(j)} ${\theta}$ = 75$^{\circ}$ and \textbf{(l)} ${\theta}$ = 90$^{\circ}$, respectively. In panel of Berry curvature plots (\textbf{(a)},\textbf{(c)},\textbf{(e)},\textbf{(g)},\textbf{(i)},\textbf{(k)}), dotted and dashed lines correspond to Fermi surfaces of lowest and second lowest bands of Cr ${e}_{\mathrm{g}}$ bands, respectively. High symmetry points are marked in panel \textbf{(a)}. Data in all panels is for one electron doping per unit cell and azimuthal angle ${\phi}$ = 0$^{\circ}$.}
	\label{fig:BC_BZ_angle}
\end{figure*}

Since spins are magnetic dipoles, magnetic long-range dipole-dipole interactions between local moments may change magnetic anisotropy. Because these inter-dipole interactions are not captured within DFT, we employ Ewald's lattice summation technique to compute dipolar energy and estimate its effect on magnetic anisotropy\cite{ewald1921evaluation,Guo_1991,PhysRevB.76.094413,PhysRevB.98.125416}. Table~\ref{table:MAE} lists dipolar interaction and anisotropy energies from DFT calculations as a function of electron doping. It is shown that dipolar interactions favor the easy-plane configuration over the easy-axis one in FM states (as shown in the negative values of D-MAE in Table~\ref{table:MAE}). Combining anisotropy energies from DFT (MAE) and dipolar interactions (D-MAE), it can be seen that easy-plane spin configurations are more favored except for undoped and one electron doping conditions. Interestingly, total magnetic anisotropy energy (MAE + D-MAE) at $\Delta n=1$ is reduced to 0.289 meV/f.u., equivalent to 1.18 Tesla of magnetic field strength. Hence controlling spin alignment and the resulting topological properties of one-electron-doped (per formula unit cell) ferromagnetic monolayer CrSiTe$_{ 3 }$ via applying an external magnetic field become achievable, which will be discussed further in the following section (Sec.~\ref{sub:in-plane magnetism}).

To investigate more details about the suppression of magnetic anisotropy energy under electron doping as depicted in Fig.~\ref{fig:band-ahc}(c), we calculate total energy changes induced by varying spin directions as illustrated in Fig.~\ref{fig:AHC_MAE_theta}(a). Without the change in the electronic structure ({\it i.e.} when the local spin moment picture is robust), total energy as a function of spin directions should have a form of (\textbf{M}${\cdot}$\textbf{z})$^{2}$ ${\sim}$ cos$^{2}{\theta}$ (see Fig.~\ref{fig:AHC_MAE_theta}(b) for the angle definition). Such trend is maintained until ${\theta}$ is increased up to 30$^{\circ}$. Beyond that angle, the anisotropy energy deviates from the simple local moment picture. The result implies an onset of additional terms that comes into play to favor the easy-plane configuration around $\theta$ = 30$^{\circ}$, which turns out to originate from the evolution of the electronic structure for $\theta$ (see below the discussion on Fig.~\ref{fig:BC_BZ_angle} for more detailed discussion). As a result, shape of total energy shows local minimum near ${\theta}$ ${\sim}$ 50$^{\circ}$ and magnitude of magnetic anisotropy energy is suppressed about 1 meV at ${\theta}$ = 90$^{\circ}$ in Fig.~\ref{fig:AHC_MAE_theta}(a).

\subsection{Switching AHC via external magnetic fields}
\label{sub:in-plane magnetism}

Because the magnetic anisotropy energy of Cr ${e}_{\mathrm{g}}$ bands in the presence of one electron doping conditions is small enough to tune the FM via external fields, we investigate the behavior of AHC as the spin orientation is changed from the out-of-plane to in-plane direction. Fig.~\ref{fig:AHC_MAE_theta}(b) shows the magnitude of AHC, averaged over the azimuthal angle $\phi$ of the net magnetization, as a function of polar angle $\theta$ of the FM spin orientation to the layer-normal direction. From ${\theta}$ = 0$^{\circ}$ to ${\theta}$ = 45$^{\circ}$, AHC remains quantized at 2${e}^2/h$ where there are deviations with respect to azimuthal spin angle ${\phi}$. Surprisingly, as ${\theta}$ increases, AHC begins to reduce and goes to almost zero at ${\theta}$ = 90$^{\circ}$. We also confirm the anti-symmetric behavior of AHC as a function of spin polar angle ${\theta}$. One can imagine that spin directions for ${\theta}$ = 180$^{\circ}$ is opposite to ${\theta}$ = 0$^{\circ}$ so that rotation of corresponding chiral edge modes are reversed, i.e. clockwise to counterclockwise. Because the direction of the FM moment can be switched between the out-of-plane and in-plane direction via external magnetic fields at one-electron doping per formula unit cell, the quantum anomalous Hall phase at this doping can also be switched on-off via external magnetic fields of about 1.18 Tesla estimated in Section~\ref{sub:dipole}. 

To understand the changes of band features and AHC at one-electron doping as the net magnetization is tilted, we plot Berry curvatures in the BZ and along with the $e_{\mathrm{g}}$ band dispersion with tilting the spin orientation direction as summarized in Fig.~\ref{fig:BC_BZ_angle}. Here we set spin azimuthal angle $\phi=0$ to plot Berry curvatures and band structures. As ${\theta}$ increases, electron and hole pockets start to develop, close to $\theta=45^\circ$, in the middle of the $\Gamma$-$M_{1,2,3}$ lines and around $M_{1,3}$ points (see Fig.~\ref{fig:BC_BZ_angle}(e) and (f)). Especially, the presence of hole pockets and their expansion contribute to the reduction of the AHC as ${\theta}$ increases beyond 30$\sim$45$^\circ$ (compare with Fig.~\ref{fig:AHC_MAE_theta}(b)), while the AHC contributions from electron pockets around M$_{1,3}$ points are almost vanishing. 

As $\theta$ is further increased beyond 60$^\circ$, sign of Berry curvature distribution around $M_{1}$ and $M_{3}$ points are flipped (compare Fig.~\ref{fig:BC_BZ_angle}(g) with (i) and (k)). This behavior is attributed to the band touching below the Fermi level and the resulting sign reversal of Berry curvature of occupied bands. Comparing panel Fig.~\ref{fig:BC_BZ_angle}(h), (j), and (l), it can be noticed that band crossings occur on $\Gamma$-$M_{1}$ and $\Gamma$-$M_{3}$ lines, slightly below the Fermi level, around $\theta$ = 75$^\circ$. The crossings give rise to sign flipping of the Berry curvature of involved bands in the vicinity of the crossing point, which leads to cancellation of net Berry curvature and vanishing AHC at $\theta$ = 90$^\circ$.

In addition, note that the band gap between highest-occupied and lowest-unoccupied bands at $\Gamma$ is suppressed as $\theta$ is increased, so that at $\theta$ = $90^\circ$ the quadratic band touching is restored. This occurs because {\it i)} the bands close to the $\Gamma$ point consist mostly of Te $p_{x,y}$-orbitals, and {\it ii)} SOC in the presence of FM behaves as an orbital Zeeman fields $\lambda_{\rm SO} {\bf m} \cdot \hat{\bf L}$ ($\lambda_{\rm SO}$ and ${\bf m}$ being SOC strength and net magnetization, respectively). When ${\bf m} = m_z {\bf z}$ the $\lambda_{\rm SO} m_z \hat{L}_z$ splits the $p_{x,y}$ doublet into $\hat{L}_z$ eigenstates ($p_x \pm i p_y$), but when ${\bf m}$ is in-plane the splitting cannot occur due to the absence of $p_{z}$ character. Hence tilting the spin direction leads to the quenching of SOC in the vicinity of $\Gamma$ point. 

\subsection{Tight binding model}
\label{sec:tbmodel}

To explore possibilities of further tuning the topological nature of this compound and obtain a deeper understanding of high Chern numbers, we construct a model Hamiltonian that correctly captures characteristics of the conduction bands. As seen in the Fig.~\ref{fig:band-pdos}, the four conduction bands are composed in Cr-$e_g$ and Te-$p$ orbitals. We can integrate out the Te-$p$ orbitals and construct the $e_g$ effective Hamiltonian as follow:
\begin{align}
    H = \sum_{i,j,\alpha,\beta} t_{i\alpha,j\beta}d_{i,\alpha}^{\dagger}d_{j,\beta}+ i t_{i\alpha,j\beta}^{soc}d_{i,\alpha}^{\dagger}d_{j,\beta},
\end{align}
where the first and second term represent the spin-independent and and spin-dependent hopping, respectively. The detailed derivation can be found in the Supplementary information section 3.

\begin{figure}
	\includegraphics[width=0.5\textwidth]{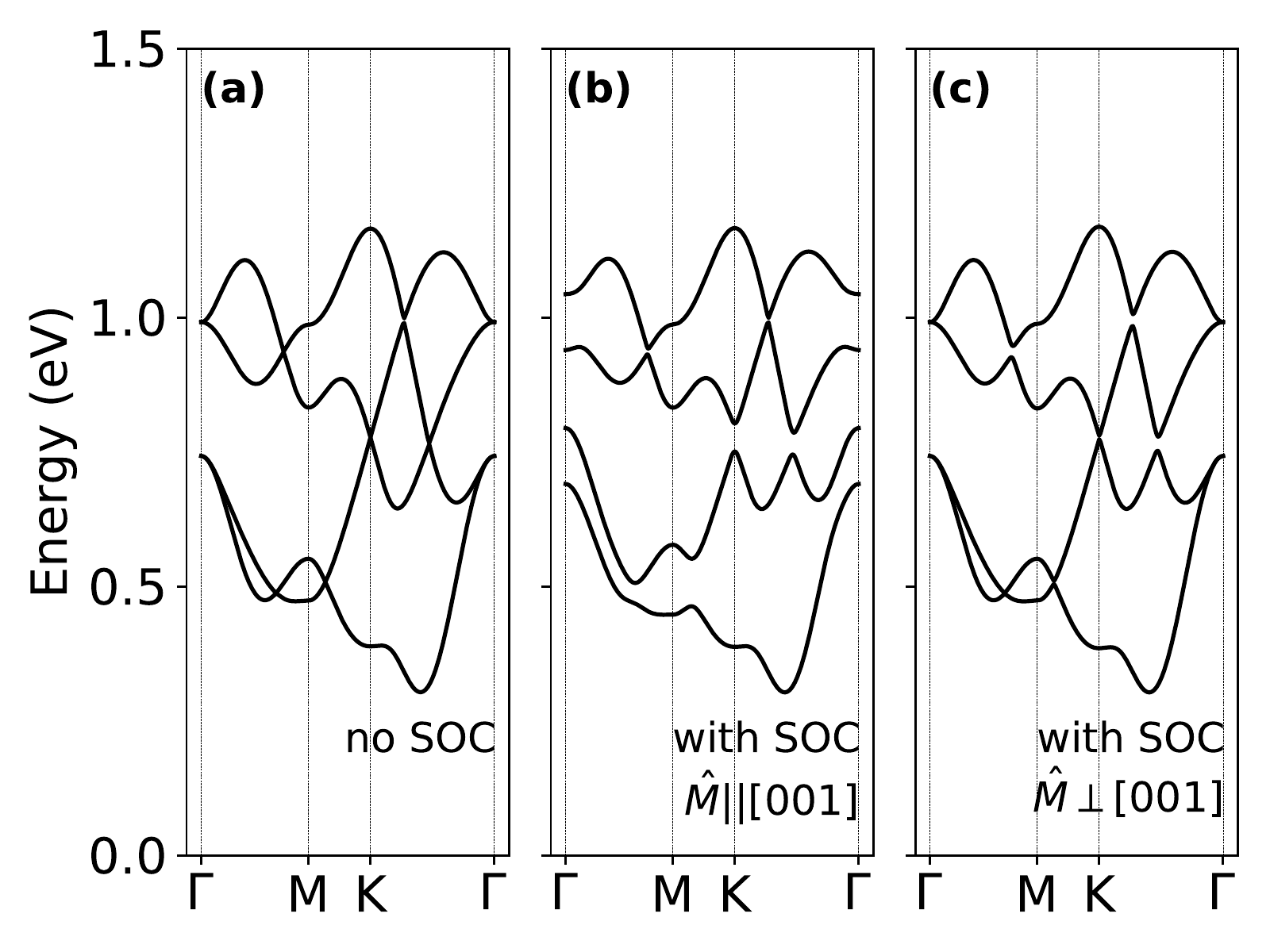}
	\caption{\textbf{Band structures from the tight binding model on four conduction bands above the Fermi level.} \textbf{(a)} without SOC effect \textbf{(b)} with SOC effect and magnetization along [001] direction, and \textbf{(c)} with SOC effect and magnetization direction along [100] direction.}
	\label{fig:tb-bands}
\end{figure}

Fig.~\ref{fig:tb-bands} shows the TB band structure without SOC effect. The hopping parameters are obtained from the Wannier function analysis. Since the TB model is constructed in the perfect octahedron system without trigonal distortion for simplicity, there exists a slight difference in the energy dispersion compared to the DFT conduction band in Fig.~\ref{fig:band-ahc}, but the overall features agree well. When we include the $t^{soc}$ parameters in the model, the bandgap opens, and each band shows [-2, 6, 8, -4] of Chern number, which also verifies the validity of the model Hamiltonian.

In this model, the SOC-dependent Hamiltonian has quiet an interesting form. For instance, the NN SOC Hamiltonian ($H_1^{soc}$) simply looks as
\begin{align}
\label{eq:tb-spin-dependent}
H_1^{soc} &=
\begin{pmatrix}
  0 & 0 & 0 & 1 \\
  0 & 0 &-1 & 0 \\
  0 & 1 & 0 & 0 \\
 -1 & 0 & 0 & 0
\end{pmatrix}
\cdot it^{soc}_{1}(\textbf{k}) \\
 t^{soc}_{1}(\textbf{k}) &= t_1^{soc}\hat{M} \cdot( \hat{z} + \hat{x} e^{-i2\pi k_1} + \hat{y} e^{-i2\pi k_2}),
\end{align}
where $\hat{M}$ denotes the unit vector for magnetization direction. In this equation, each vector $\hat{x}$, $\hat{y}$, and $\hat{z}$ represents the local coordinates denoted in Fig.~\ref{fig:structure}. This magnetization-dependent hopping gives a different effect on each $k$ point. Since the gap closing is observed at some high symmetry $k$ points, let us examine how these parameter changes on each $k$ point.

At $\Gamma$ point, the parameter is proportional to $\hat{M} \cdot \vec{c}$, where $\vec{c}$ denotes the [001] direction. Thus, the SOC effect becomes maximum when the spins point out-of-plane direction, while it vanishes when the spin lies on the plane. However, we have slightly different behavior on $K=(2/3,1/3)$ point. In this case, the matrix element vanishes when the magnetization points out-of-plane direction ($t_1^{soc}(K) \sim \hat{c} \cdot(\hat{z} + \hat{x}e^{-i4\pi/3}+\hat{y}e^{-i2\pi/3})=0$. For the general $k$ points, this term does not vanishes. However, due to the dependence on the local axis, these terms cancel each other and become relatively small. Thus, we can observe that the band structure from the full SOC Hamiltonian when the magnetization lies in-plane looks very similar to the FM band structure without SOC effect. 

\begin{figure}
    \includegraphics[width=\linewidth]{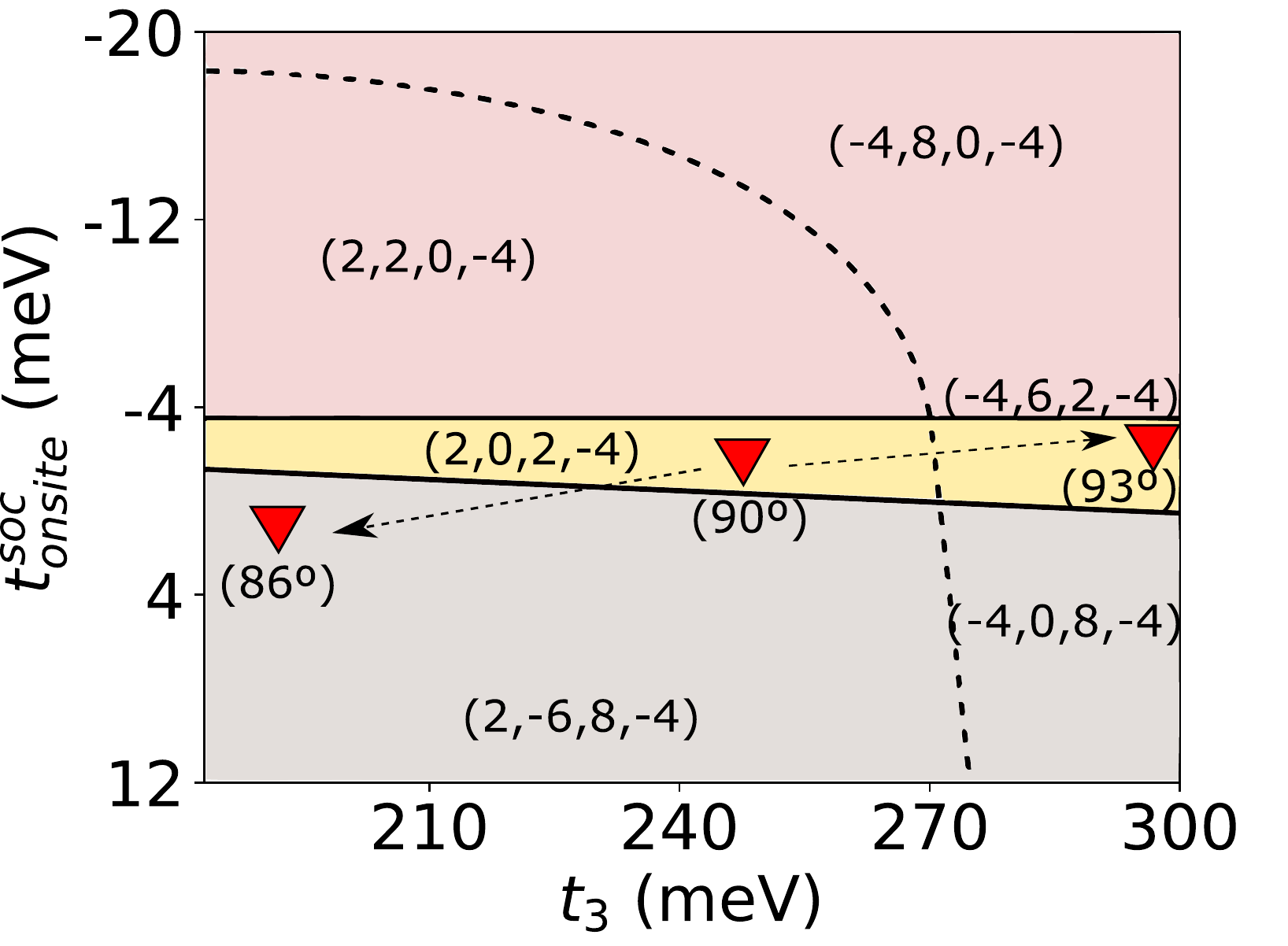}
    \caption{\label{fig:chern-phasel} \textbf{Change of the Chern number of each band with respect to hopping parameters in tight binding model}. $t_3$ and $t_{onsite}^{soc}$ represent $3^{rd}$ NN hopping and onsite SOC term, respectively. The four integer pair ($i_1$, $i_2$, $i_3$, $i_4$) represent the Chern number of each band. DFT calculations with three different structures which have distinct $\textrm{Cr-Te-Cr}$ angles(X$^{\circ}$) was also performed, and the corresponding parameter values extracted from the Wannier function analysis are makerd with red triangle.
} 
\end{figure}

To examine the relation between the parameter strength and topological character, we also examined the change of Chern number of each band by varying some parameters. Fig.~\ref{fig:chern-phasel} shows the evolution of the Chern number while varying two parameters $t_3$($3^{rd}$ NN hopping) and $t_{onsite}^{soc}$(onsite SOC term). Other parameters such as NN and $2^{nd}$ NN terms are extracted from the Wannier function analysis. In this figure, the four integer pair ($i_1$, $i_2$, $i_3$, $i_4$) represent the Chern number of each band calculated with model Hamiltonian. We can observe that the Chern number set of (2,$-$6,8,$-$4), the actual Chern number of the CrSiTe$_3$ system, is observed on the lower-left parameter region. With two-electron doping, the total Chern number equals $-$4, which does not change even if we change $t_3$. It change from 4 to 2 with $t_{onsite}^{soc}$ being 0 $\sim$ $-$4 meV.

Interestingly, this figure shows that structure distortion can achieve in each phase. The fully optimized structure of CrSiTe$_3$ suffers trigonal distortion. The angle between Cr-Te-Cr is about $86^{\circ}$.
In this case, $t_3$ and $t_{onsite}^{soc}$ becomes about 180 and -1 meV as marked with red triangle in the figure. If we perform the DFT calculation in the cubic structure where the local octahedron has no distortion(Cr-Te-Cr angle being $90^{\circ}$), the Chern number changes from (2,$-$6,8,$-$4) to (2,0,2,$-$4). The hopping parameters corresponding to this case is about 240 and $-$3 meV for $t_3$ and $t_{onsite}^{soc}$, respectively.

To determine the effect of structure distortion on the change of the Chern number, we performed the DFT calculations on the various structures where Cr-Te-Cr angles were set to be between $86$ (fully optimized structure) and $90^{\circ}$ (no distortion). On each step, we extracted $t_3$ and $t_{onsite}^{soc}$ parameters. Other parameters also vary, but their changes were minimal compared to these two parameters, so we fixed them. Each parameter increased linearly from the fully optimized structure marked with a dotted line. During this transition, the total Chern number changes from $-$4 to 2. Even though it is not an optimized structure, we can further distort the local octahedron to achieve the Cr-Te-Cr angle to be $93^{\circ}$ and monitor the change of hopping parameters. In this case, the Chern number becomes ($-$4,6,2,$-$4), and the corresponding $3^{rd}$ NN hopping parameter becomes about 300 meV. During this process, the Chern number changes when Cr-Te-Cr angles become about $91.8^{\circ}$ ($t_3$ being about 270 meV). These observation implies that the local distortion near metal site (Cr) on CrSiTe$_3$ induces large modification of $t_3$ and $t_{onsite}^{soc}$ parameters and phase transition. The significant modification of $t_3$ comes from the modification of bond length between Te $p$ orbitals. 

\begin{table}[h]
\renewcommand{\arraystretch}{1.2}
\renewcommand{\tabcolsep}{2.0mm}
\begin{tabular}{cccc}
\hline
\hline
& Fully & No & Trigonal\\ 
& Optimized & Distortion & Compression\\
\hline
$\angle \textrm{Cr-I-Cr}$($^\circ$)& 86 & 90 & 93 \\
Te-Te(\AA) & 4.23 & 4.04 & 3.90 \\
Cr-Te(\AA) & 2.95 & 2.86 & 2.80 \\
$\overline{t_1}$ (meV) & 30 & 44 & 67 \\
$\overline{t_2}$ (meV) & 18 & 16 & 15 \\
$\overline{t_3}$ (meV) & 176 & 249 & 308 \\
$\overline{t_5}$ (meV) & 40 & 38 & 44 \\
$\overline{t_{onsite}^{soc}}$ (meV) & -0.8 & -2.2 & -3 \\
$\overline{t_{1}^{soc}}$ (meV) & 14 & 10 & 7 \\
$\overline{t_{3}^{soc}}$ (meV) & 16 & 21 & 24 \\
$\overline{t_{2}^{soc}}$ (meV) & 4 & 5 & 8 \\
\hline
\hline
\end{tabular}
\caption{Structure and hopping parameters of fully optimized, cubic, trigonally compressed structure. The distortion are only applied to local octahedron, and the lattice constant remains the same. $\overline{t_i}$ represents the averaged transfer integral $\overline{t_i} = \sqrt{\textrm{Tr}({H_{0i}H_{i0}})}$ between the origin(0) and the $i$th neighbor site.}
\label{table:bonding-angle}
\end{table}

Table~\ref{table:bonding-angle} summarizes the structure information on each calculation. We can see that the Te-Te bond length decreases from 4.23 (fully optimized) to 3.90 \AA\ (trigonally compressed distortion). As discussed, the $3^{rd}$ NN hopping contains Te $\rightarrow$ Te hopping channel. This considerable reduction of bond length between Te sites increases the $t_3$ parameter. We can also observe that the Cr-Te distance decreases. This parameter is related to $t_{onsite}^{soc}$, and we also observe that this value slightly increases. Other parameters such as $t_2$ or $t_5$ remains almost the same.  The local distortion of the octahedron can be achieved by applying external strain, and it induces a change in the hopping parameters. Thus, we can effectively map the effect of distortion to the change of hopping parameters, although this Hamiltonian assumes a cubic structure. Since the Chern number depends on the hopping parameters and the optimized structure lies near the phase boundary, we expect it to be possible to tune the Chern number phase by tuning external strain or change of ligand hopping strength.

\section{DISCUSSION}
\label{sec:discussion}

We study the electronic and topological properties of the electron-doped single-layer structure of CrSiTe$_{3}$ by performing first-principle calculations based on density functional theory (DFT). The lattice constant and Hubbard $U$ parameters are fixed by confirming that FM ground state and topological characters are invariant under certain conditions. We construct MLWF, the Fourier transform of Kohn-Sham wavefunctions of each band, and calculate Berry curvatures to analyze topological characters using converged MLWF. Nontrivial topology appears within Cr $e_{\mathrm{g}}$ band manifold which can lead to Chern insulating phases in CrSiTe$_{3}$. Chiral edge modes are also calculated, consistent with total Chern number calculations. Moreover, we find spin angle-dependent AHC under one-electron doping in the formula unit cell and suppression of magnetic anisotropy. For our calculation, one-electron doping in the formula unit cell with 7.0{\AA} lattice constant corresponds to carrier density as 2.36$\times$10$^{14}$ cm$^{-2}$. Ionic liquid gating is representative method of charge doping where order of doping concentration is known as 10$^{14}$ cm$^{-2}$ in 2D systems \cite{adma201607054,mi12121576,doi:10.1021/nn401053g}. In recent study, about 2$\times$10$^{14}$ cm$^{-2}$ of electron doping for one layer is accomplished in CrGeTe$_{3}$\cite{nelecton2520} almost adjacent to our theoretical goal. Because CrGeTe$_{3}$ has the same crystal structure and similar lattice constant as CrSiTe$_{3}$, we expect our theoretical suggestion of Chern insulating phases and magnetic field dependent QAHE can be realized as shown throughout section~\ref{sec:results}. By performing tight-binding (TB) model analysis, we confirm that DFT calculation results of Chern number for Cr $e_{\mathrm{g}}$ orbitals are reproduced. We also find distinct Chern insulating phases by adjusting hopping parameters in our constructed TB model and drawing relevant phase diagrams. We expect that CrSiTe$_{3}$ can be a Chern insulator controlled by the external magnetic field, which can be realized by electron doping.

\section*{METHODS}
\label{sec:methods}
\subsection*{Density Functional Theory Calculations}
\label{sub:dft-calc}

To obtain band structures and projected density-of-states (pDOS), we perform ${ab\ initio}$ electronic structure calculations based on density-functional theory (DFT) using {\sc OpenMX}\cite{openmx,PhysRevB.72.045121} code, which employs linear-combination-of-pseudo-atomic-orbital basis with norm-conserving pseudopotentials. We choose generalized gradient approximation (GGA) exchange-correlation functional in the parameterization of Perdew, Burke and Enzerhof (PBE)\cite{PhysRevLett.77.3865} with Hubbard $U$ parameters chosen to be 1.5 eV for Cr ${d}$ orbital. SOC effects are included in the calculations via fully-relativistic pseudopotentials implemented in {\sc OpenMX}\cite{openmx,PhysRevB.72.045121}. Pseudo-atomic orbitals are set to be ${s}^{\mathrm{3}}$${p}^{\mathrm{2}}$${d}^{\mathrm{2}}$ for Cr, ${s}^{\mathrm{2}}$${p}^{\mathrm{2}}$${d}^{\mathrm{1}}$ for Si, ${s}^{\mathrm{3}}$${p}^{\mathrm{3}}$${d}^{\mathrm{2}}$ for Te, respectively. To simulate two-dimensionality, we insert 20 \AA~of vacuum in the unit cell. 10 $\times$ 10 $\times$ 1 of $k$-space mesh is adopted for the momentum space integration. Energy cutoff for choosing real-space grid is set to be 700 Ry (96 $\times$ 96 $\times$ 360 real space grid). 10$^{-6}$ Hartree/Bohr of force criterion is chosen for the optimization of internal coordinates while keeping ${C}_{\mathrm{3}}$ rotation, inversion and mirror symmetries. SOC effects are excluded in the process of structural relaxation. We fix Bravais lattice as hexagonal and determine lattice constant by performing total energy minimization calculation as a function of unit cell size. 

\subsection*{Analysis of topological characteristics}
\label{sub:analysis}
Berry curvature is given by
\begin{equation}
\label{eq:berry}
    B_{n}(\textbf{k}) = i\sum_{n^{\prime}{\neq}n} {{\langle} n {\mid} {{\partial}H \over {\partial}k_{x}} {\mid} n^{\prime} {\rangle}{\langle} n^{\prime} {\mid} {{\partial}H \over {\partial}k_{y}} {\mid} n {\rangle} - (k_{x} {\leftrightarrow} k_{y}) \over (E_{n} - E_{n^{\prime}})^2}
\end{equation}
where integrating Berry curvature in BZ gives Chern number
\begin{equation}
\label{eq:chern}
    C_{n} = {1 \over 2\pi} {\iint_{BZ}}B_{n}(\textbf{k}) dk_{x}dk_{y}
\end{equation}
$E_{n}$ is Kohn-Sham eigen energy of band index $n$ at certain $\textbf{k}$=($k_{x}$,$k_{y}$) point and $H$ is Hamiltonian of system. Anomalous Hall conductivity is calculated by integrating Berry curvatures over BZ up to arbitrary energy level which is given by
\begin{equation}
\label{eq:AHC}
    \sigma_{H}(E) = {e^2 \over h} {\iint_{BZ}}B_{n}(\textbf{k})f_{FD}(E,\textbf{k}) dk_{x}dk_{y}
\end{equation}
where $e$ is electron charge, $h$ is Planck constant and $f_{FD}(E,\textbf{k})$ is Fermi-Dirac distribution function, respectively.
To compute Berry curvatures, anomalous Hall conductivity, Chern numbers, and chiral edge states, the maximally localized Wannier function (MLWF) method as implemented in {\sc OpenMX} is employed\cite{RevModPhys.84.1419,PhysRevB.79.235118}. ${d}_{\mathrm{z}^{2}}$ and ${d}_{\mathrm{x}^{2}-{y}^{2}}$ orbitals at Cr sites are chosen as initial projectors. 10 $\times$ 10 $\times$ 1 $k$-space grid is chosen for the construction of MLWFs. From the MLWF tight-binding Hamiltonian obtained, {\sc Wannier Tools} code is used to compute Berry curvature, Chern numbers, and edge states\cite{wu2018wanniertools}. Specifically, we employ Fukui-Hatsugai formalism\cite{Fukui2005} to compute Berry curvature. Moreover, edge state calculations utilize the iterative Green's functions method in the semi-infinite geometry.

\subsection*{Generalized tight-binding model for $e_{\mathrm{g}}$ manifold}
\label{sub:tb-model}
In addition to the MLWF tight-binding model, we further introduce a Slater-Koster-type tight-binding model for a deeper understanding of SOC and lattice distortion effects. 

We first start from including both Te $p$ orbitals and Cr $d$ orbitals, then project out Te $p$ orbitals to obtain an effective four-band Hamiltonian. 

\section*{DATA AVAILABILITY}
All data are available from the corresponding author upon reasonable request. \\

\section*{ACKNOWLEDGMENTS}
	We gratefully acknowledge Kee Hoon Kim for valuable discussions. This work was supported by the National Research Foundation of Korea (NRF) (no. 2020R1F1A1066548). In addition, H-SK acknowledges the support of the National Research Foundation of Korea (Basic Science Research Program, Grant No. 2020R1C1C1005900).

\renewcommand{\refname}{\textsc{REFERENCES}}
\bibliography{CrSiTe3-npj}




\newpage

\section*{Supplementary information: GRAPHIC OF MLWF AND SIMPLE TIGHT BINDING MODEL APPROACH}
\label{appendix:tbcomponents}

\begin{figure}
  \includegraphics[width=1.0\linewidth]{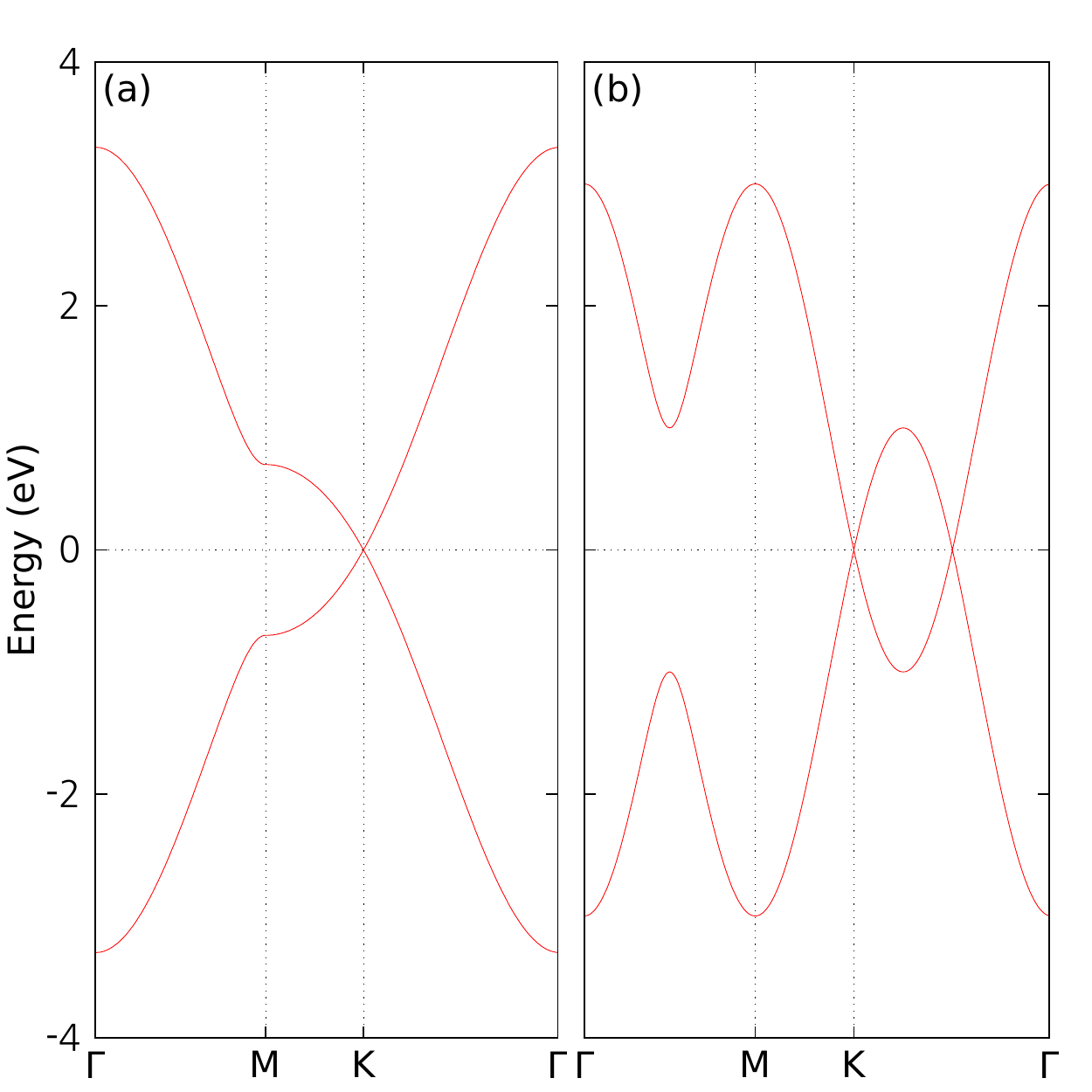}
  \caption{\textbf{Band structure plot of simple $s$ orbital like two band tight-binding model} \textbf{a} Dominant 1$^{st}$ NN hoping ($t_1$ $>>$ $t_3$) and \textbf{b} dominant 3$^{rd}$ NN hoping ($t_1$ $<<$ $t_3$) conditions. The 2$^{nd}$ NN hopping term is set to be zero because it only affects energy shift of band diagram where overall features of band dispersion are remain unchanged.}
  \label{fig:toytb}
\end{figure}

We first start with a graphene-like case for a simple tight-binding approach as a toy model. If we consider $s$ band like orbital with NN hopping in a honeycomb lattice, the Dirac cone appears at the $K$ point. When 2$^{nd}$ and 3$^{rd}$ NN hopping are taken into account, the Hamiltonian is given by following expressions which becomes more complicated
\begin{align}
\label{eq:toytb}
H(\textbf{k}) &=
\begin{pmatrix}
  g(\textbf{k}) & f(\textbf{k})\\
  f^*(\textbf{k}) & g(\textbf{k})\\
\end{pmatrix}
.
\end{align}
Component of above Hamiltonian $f(\textbf{k})$ and $g(\textbf{k})$ is written by
\begin{equation}
\label{eq:hamcom}
\begin{split}
\\f(\textbf{k}) &= t_1[e^{ik_ya / \sqrt{3}} + 2e^{-ik_ya / 2\sqrt{3}}cos({k_xa / 2})]
\\&+ t_3e^{ik_ya / \sqrt{3}}(2cos(k_xa) + e^{-\sqrt{3}ik_ya})
\\g(\textbf{k}) &= t_2[2cos(k_xa) + 4cos({k_xa / 2})cos({\sqrt{3}k_ya / 2})],
\end{split}
\end{equation}
where $\textbf{k} = (k_x, k_y)$, $a$ is lattice constant and $t_i$ is $i^{th}$ NN hopping parameters. Diagonalizing the Hamiltonian expressed in Eq.~\ref{eq:toytb}, one can obtain equation of eigenenergy function as follows.
\begin{equation}
\label{eq:toyene}
\begin{split}
E(\textbf{k}) &= t_2(2cos(k_xa) + 4cos(k_xa / 2)cos(\sqrt{3}k_ya / 2))
\\& \pm \big[(t_1 + 2 t_3 cos(k_xa)^2 + 2t_3cos(\sqrt{3}k_ya)) + {t_3}^2 \\ &+ (t_1 + 2t_3cos(k_xa))(4t_1cos(k_xa / 2)cos(\sqrt{3}k_ya / 2) \\ & + (2t_1cos(k_xa / 2))^2 + 4t_1t_3cos(k_xa / 2)cos(\sqrt{3}k_ya / 2)\big]^{1/2}.
\end{split}
\end{equation}
In this case, additional crossing point appears near the midpoint of $\Gamma$ - $K$ line under the condition of large 3$^{rd}$ NN hopping parameter compared to 1$^{st}$ and 2$^{nd}$ NN hopping terms. This features are depicted in fig.~\ref{fig:toytb}.

\begin{figure}
  \includegraphics[width=1.0\linewidth]{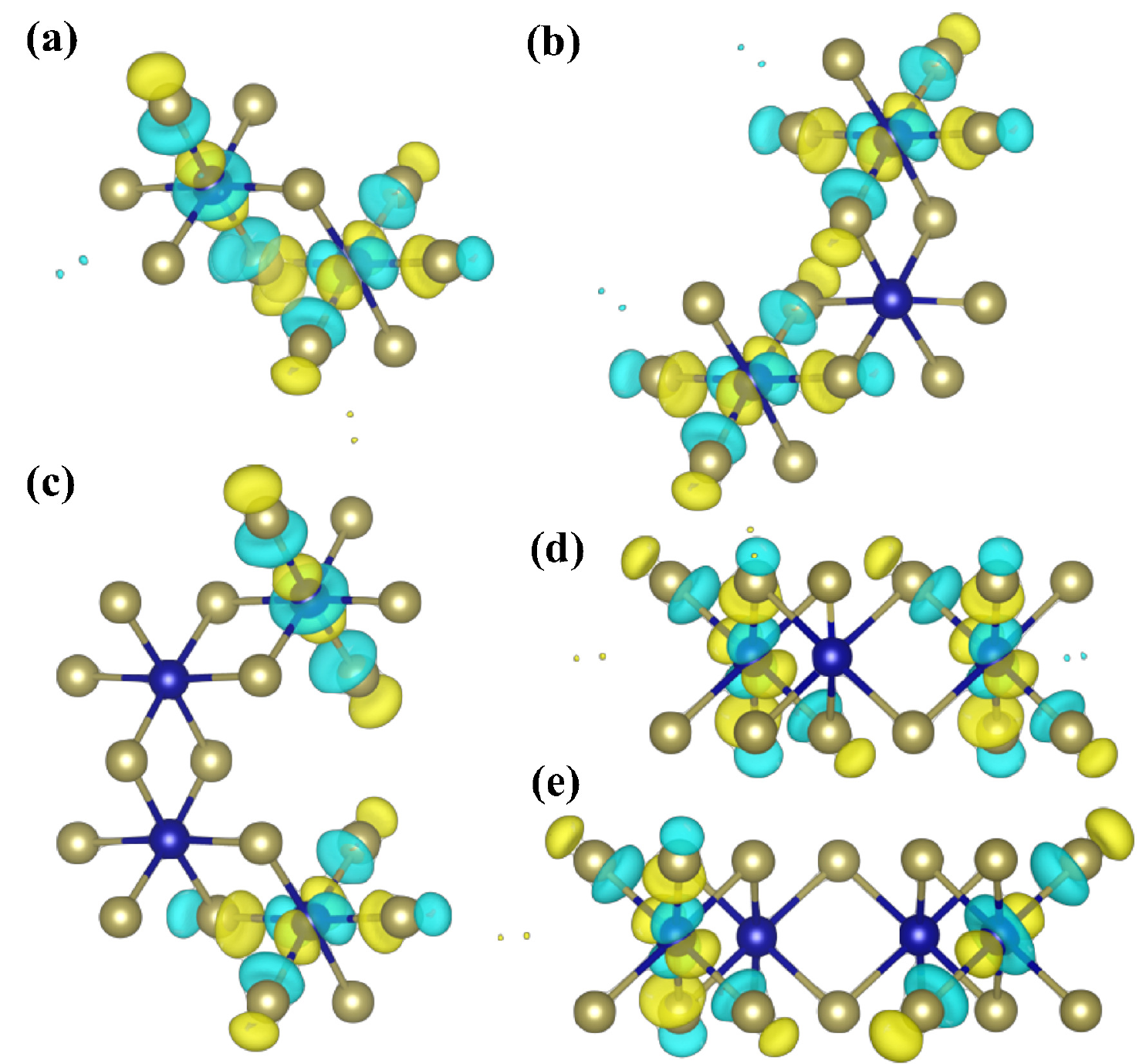}
  \caption{\textbf{Graphic of converged MLWFs} \textbf{a} top view of Nearest neighbor(NN), \textbf{b} 2$^{nd}$ NN, \textbf{c} 3$^{rd}$ NN and \textbf{d} side view of 2$^{nd}$ NN, \textbf{e} 3$^{rd}$ NN, respectively. MLWFs are constructed by combining four Cr $e_{\mathrm{g}}$ bands hybridized with Te $p$ orbitals.}
  \label{fig:WF}
\end{figure}

\begin{table}[h]
\renewcommand{\arraystretch}{1.2}
\renewcommand{\tabcolsep}{1.5mm}
\centering
\begin{tabular}{c c c c c c}
 \hline
 (${i}$, ${j}$) & ${t^{(1)}}_{ij}$ & ${t^{(2)}}_{ij}$ & ${t^{(3)}}_{ij}$ & ${t^{(4)}}_{ij}$ & ${t^{(5)}}_{ij}$\\
 \hline\hline
 $d_{x^2-y^2}$, $d_{x^2-y^2}$ & -27.97 & 6.7 & 175.08 & -2.42 & 36.94\\
 $d_{x^2-y^2}$,  $d_{z^2}$ & 2.55 & 14.46 & 10.21 & 1.76 & 0.64\\
 $d_{z^2}$,  $d_{x^2-y^2}$ & 1.51 & 13.7 & 3.65 & 1.77 & 1.09\\
 $d_{z^2}$, $d_{z^2}$ & -8.82 & -18.03 & -20.36 & 12.49 & -14.85\\
 \hline
\end{tabular}
\caption{Hopping parameters based on converged MLWF basis. ${t^{(n)}}_{ij}$ = ${\langle}$ i, 0 ${\mid}$ $\hat{H}$ ${\mid}$ j, $\vec{r_n}$ ${\rangle}$, where i, j index indicates pair of MLWFs and $\vec{r_n}$ is cell displacement vector which corresponds to n$^{th}$ NN site. Unit of hopping parameters is meV.}
\label{table:WF}
\end{table}

Table~\ref{table:WF} shows the overlap matrix from our MLWF calculation result, which corresponds to hopping integrals in TB model analysis. One can find a significant 3$^{rd}$ NN term compared to other terms which satisfy the condition of the above simple model approach. It can be confirmed qualitatively in Fig. \ref{fig:WF} which shows converged MLWF graphically. Hopping between NN sites is small due to the orthogonality of mediating Te $p$ orbitals. For 2$^{nd}$ and 3$^{rd}$ NN hopping, mediating two Te $p$ orbitals are located in different layers for 2$^{nd}$ NN case while in the same layer for 3$^{rd}$ NN case, which induces much significant 3$^{rd}$ NN hopping parameters compared to 1$^{st}$ and 2$^{nd}$ ones. This is a simple two-band model with $s$-like orbital to understand the shape of band structure briefly; thus, we considered a more realistic tight-binding model considering up to 5$^{th}$ NN hopping with four band model in the tight-binding analysis part.

\section*{Supplementary information: BANDGAP CLOSING DUE TO MIRROR SYMMETRY}
\label{appendix:mirror}
\begin{figure}
	\includegraphics[width=1.0\linewidth]{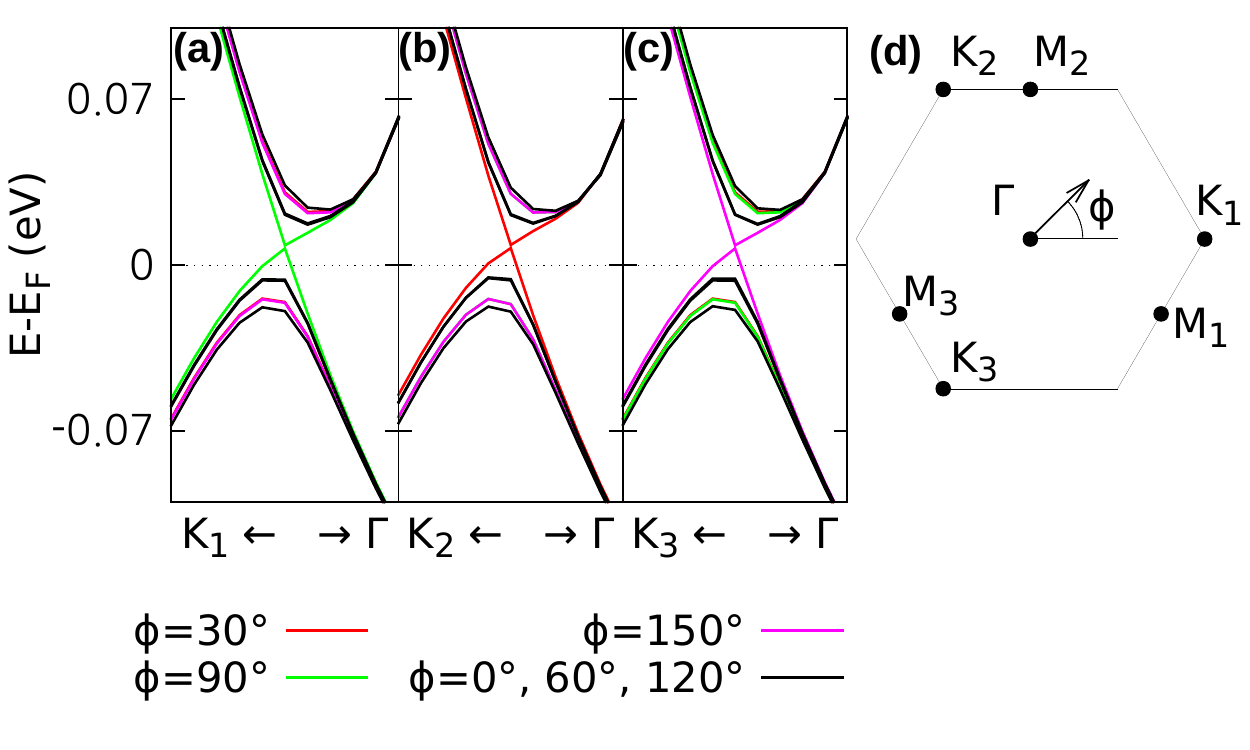}
	\caption{\textbf{Band structure with spin easy plane ($\theta$ = 90$^\circ$) under two electron doping} Expanded near middle of \textbf{a} ${\Gamma}$ - ${K_1}$, \textbf{b} ${\Gamma}$ - ${K_2}$ and \textbf{c} ${\Gamma}$ - ${K_3}$ line, respectively. \textbf{d} First BZ with high symmetry points and in-plane spin angle ${\phi}$.}
	\label{fig:in_plane_spin}
\end{figure}

If spin lies in-plane, its direction has chance to become vertical to one of three mirror planes of CrSiTe$_{ 3 }$ which contain ${\Gamma}$ - ${K}$ line and perpendicular to its layer. When spin orientation is perpendicular to mirror plane, mirror symmetry would then be restored which is shown in Fig. \ref{fig:in_plane_spin}. For case of ${\phi}$ = 90${\degree}$, 30${\degree}$ and 150${\degree}$, band gap is closing near midpoint of ${\Gamma}$ - ${K_n}$ (where n=1,2,3) line, respectively while still maintaining at other points including ${K_n}$ points. In Fig. \ref{fig:in_plane_spin}(d), one can confirm each spin angle (${\phi}$ = 30${\degree}$, 90${\degree}$, 150${\degree}$) corresponds to perpendicular to mirror plane (${\Gamma}$ - ${K_2}$, ${\Gamma}$ - ${K_1}$, ${\Gamma}$ - ${K_3}$), respectively. We may conclude additional crossing points at $\Gamma$ - $K$ line originated from mirror symmetry of the system, while crossing at K point is from sublattice symmetry which is equivalent with graphene case.

\section*{Supplementary information: Derivation of tight binding Hamiltonian}
\label{appendix:tbmodel}
The full Hamiltonian $H$ of on Cr-$d$ and Te-$p$ orbitals can be written as
\begin{align}
    H &= H_{0} + H_{CF} + H_{soc} + H_{t} + H_{\textbf{B}_\textrm{eff}} \\
    H_{CF} &= \sum\limits_{i, j=-2}^{2} V_{i,j}^{CF} d_{i}^{\dagger} d_{j} \\
    H_{soc} &= \lambda_{p} \hat{L_{p}} \cdot \hat{S}_{p} \\
    H_{0} &= \sum\limits_{\alpha, \textbf{r}} \epsilon_{d} d_{\alpha, \textbf{r}}^{\dagger} d_{\alpha, \textbf{r}} + \sum\limits_{\beta, \textbf{r}} \epsilon_{p} p_{\beta, \textbf{r}}^{\dagger} p_{\beta, \textbf{r}} \\
    H_{t} &= \sum\limits_{c, \alpha,\textbf{r}, \beta, \textbf{r}'} V(\alpha,\textbf{r}, \beta, \textbf{r}') c_{\alpha, \textbf{r}}^{\dagger} c_{\beta, \textbf{r}'} \\    
    H_{\textbf{B}_{\textrm{eff}}} &= \sum_{\textbf{r},\alpha,\tau,\tau'} [\textbf{B}_{\textrm{eff}}(\textbf{r}) \cdot \boldsymbol{\sigma}]_{\tau, \tau'} c^{\dagger}_{\alpha, \tau, \textbf{r}} c_{\alpha, \tau', \textbf{r}},    
\end{align}
where $H_{0}$, $H_{CF}$, $H_{SOC}$, $H_t$, and $H_{\textbf{B}_\textrm{eff}}$ represent the orbital energy, crystal field on $d$ orbitals, SOC, hopping, and effective Zeeman Hamiltonian, respectively. In the hopping Hamiltonian $H_t$, $V(\alpha,\textbf{r}, \beta, \textbf{r}')$ represents the Slater-Koster parameter between orbital $\alpha$ at $\textbf{r}$ and $\beta$ at $\textbf{r'}$. Since the magnitude of SOC on Te is much larger than Cr, we only include SOC Hamiltonian of $p$ orbitals. The Zeeman Hamiltonian simulates the magneitc ordering. For instance, $\textbf{B}_\textrm{eff}(\textbf{r})$ is set to be the along [001] for all metal sites in the out-of-plane magnetization. This gives the magnetization-dependent hopping terms as Eq. 2 and Eq. 3 in our main article. To simplify the model, we integrate out the $p$ orbital degree of freedom and construct the effective Hamiltonian on $e_g$ orbitals through Kato-Takahashi perturbation formalism\cite{kato1949convergence, takahashi1977half}. The explicit parameters are given as follow:

\begin{align}
    t_{1} &= V_{dd \pi}, \\
    t^{'}_1 &= \frac{3V_{dd \pi}}{4} + \frac{V_{dd \sigma}}{4} \\ 
    t_2 &= \frac{V_{pd \sigma}^2 V_{pp \pi}}{(E_d - E_p)^2} \\
    t_3 &= \frac{V_{pd \sigma}^2 (-V_{pp \pi} + V_{pp \sigma})}{4(E_d - E_p)^2} \\
    t_{5} &= \frac{V_{pds}^2}{4(E_d-E_p)^3}(V_{ppp}-V_{pps}) (V_{ddp}-V_{ppp}-V_{pps}) \\
    t^{soc}_{onsite} &= \frac{\sqrt{3} V_{pds}^{2} \lambda_{p}}{16 (Ed - Ep)^{3}} (3 V_{ddd} + 4 V_{ddp} + V_{dds}) \\ 
    t^{soc}_{1} &= \frac{\sqrt{3}V_{pds}^2 \lambda_{p}}{4(E_d - E_p)^2} \\
    t^{soc}_{3} &= \frac{\sqrt{3}V_{pds}^2 \lambda_{p}(V_{ppp}+V_{pps})}{4(E_d - E_p)^3} \\
    t^{soc}_{2} &= \frac{\sqrt{3}V_{pds}^2 \lambda_{p}(V_{ppp}-V_{pps})}{4(E_d - E_p)^3}
\end{align}

\end{document}